\documentclass[fleqn,twoside,11pt]{article}

\usepackage{amssymb}
\usepackage{graphicx}
\usepackage{epsbox}
\usepackage{amsmath}

\begin{document}
\noindent
{\sf University of Shizuoka}

\hspace*{13cm} {\large US-05-03}

\vspace{2mm}

\begin{center}

{\Large\bf  Challenge to the Mystery of }\\[.1in]
{\Large\bf the Charged Lepton Mass Formula }\footnote{
Contribution paper to Lepton-Photon 2005}

\vspace{2mm}
{\bf Yoshio Koide}

{\it Department of Physics, University of Shizuoka, 
52-1 Yada, Shizuoka 422-8526, Japan\\
E-mail address: koide@u-shizuoka-ken.ac.jp}

\end{center}

\begin{abstract}
Why the charged lepton mass formula 
$m_e +m_\mu +m_\tau = \frac{2}{3} (\sqrt{m_e}+\sqrt{m_\mu}
+\sqrt{m_\tau})^2$ is mysterious is reviewed, and guiding 
principles to solve the mystery are presented.  According 
to the principles, an example of such a mass generation 
mechanism is proposed, where the origin of the mass spectrum 
is attributed not to the structure of the Yukawa coupling
constants, but to a structure of vacuum expectation
values of flavor-triplet scalars under Z$_4 \times$S$_3$
symmetries.
\end{abstract}

\vspace{3mm}

{\large\bf 1. Introduction}

It is widely accepted that quarks and leptons 
are fundamental entities of the matter.
If it is true, the masses and mixings of the 
quarks and leptons will obey a simple law of nature, 
and we will be able to find a beautiful relation 
among those values. 
Nowadays, for all of the 3 family quarks, 
we know their mass values, but the accuracy of 
those values is still somewhat unsatisfactory 
for testing the validity of the model rigorously. 
The experimental situation in the neutrino masses and mixings 
is also not in the satisfactory accuracy. 
In contrast to quarks and neutrinos, 
for the charged leptons, 
we know their mass values with sufficient accuracy. 
If we can find a beautiful mass 
(and mixing) relation, 
it will make a breakthrough in the unified understanding 
of the quarks and leptons. 

In 1992, the observed tau lepton mass value was revised by new experiments
\cite{taumass} as
$$
m_{\tau}^{old}=1784 \pm 4 \ {\rm MeV} \ \  \Longrightarrow 
\ \ m^{new}_{\tau}=1776.99^{+0.29}_{-0.26} \ {\rm MeV}.
\eqno(1.1)
$$
(The new value has been quoted from Ref.~\cite{PDG04}.)
Since the new value $m_\tau=1777$ MeV has already been predicted
by a charged lepton mass formula \cite{Koide82,Koide83,Koide90}
$$
m_e+m_{\mu}+m_{\tau}=
\frac{2}{3}(\sqrt{m_e}+\sqrt{m_\nu}+\sqrt{m_{\tau}} )^2 , 
\eqno(1.2)
$$
the mass formula had received considerable attention at one time.
Indeed, the formula (1.2) predicts the tau lepton mass value 
$$
m_{\tau}=1776.97 \ {\rm MeV},
\eqno(1.3)
$$
from the observed electron and muon mass values 
\cite{PDG04} $m_e=0.51099892 {\rm MeV}$ and 
$m_{\mu}=105.658369$  {\rm MeV}. 
The predicted value (1.3) is in excellent agreement 
with the observed value (1.1) \cite{PDG04}.
The excellent agreement seems to be beyond a matter
of accidental coincidence,
so that we should consider the origin of the mass 
formula (1.2) seriously.
However, up to the present, the theoretical basis of 
the mass formula (1.2) is still not clear.


The formula was first found \cite{Koide82} in 1982
on the basis of a composite model of quarks and leptons. 
Here,
we have assumed that the charged lepton masses $m_{ei}$ are
described as 
$$
m_{ei}=m_0(z_i + z_0)^2 , 
\eqno(1.4)
$$
where
$$
z_1 + z_2 + z_3=0 ,
\eqno(1.5)
$$
$$
z_0=\frac{1}{\sqrt3}\sqrt{z^2_1 + z^2_2 + z^2_3}.
\eqno(1.6)
$$
However, such the scenario based on a composite model 
has not been justified from the field theoretical point of view.
The explicit expression (1.2) was given in Ref.~\cite{Koide90}.
Here, a mixing between octet and singlet states in the U(3) family 
symmetry has been assumed.
Since 1993, several authors \cite{Foot,Esposito,Li-Ma} have challenged to give 
an explanation of the mass formula (1.2), 
but, at present, there is no convinced one. 

\vspace{5mm}
{\large\bf 2. How the formula is mysterious}

The charged lepton mass formula (1.2) has 
the following peculiar features:

\noindent(a) The mass formula is described in terms of the 
root squared masse $\sqrt{m_{ei}}$.

\noindent(b) The mass formula is invariant under the exchanges 
$\sqrt{m_{ei}} \leftrightarrow \sqrt{m_{ej}}$.
We know that the electron mass 
$m_e$ is negligibly small 
compared with other charged lepton masses. 
If we put $m_e=0$ in the formula (1.2), 
we will obtain a wrong prediction 
$m_{\tau}=[(\sqrt3 +1)/(\sqrt3 -1)]^2 m_\mu =1471.63 \ {\rm MeV}$
 instead of (1.3).
Thus, the non-zero value of $m_e$ is essential in the formula (1.2).

\noindent(c) The formula gives a relation between mass ratios 
$\sqrt{m_e / m_{\mu}}$ and $\sqrt{m_{\mu} / m_{\tau}}$, 
whose 
behaviors under the renormalization group equation (RGE) 
effects are different from each other. 
Therefore, the formula (1.2) is not invariant 
under the RGE effects. 
The formula is well satisfied at a low energy scale 
rather than at a high energy scale. 

\vspace{2mm}
\unitlength=1cm
\hspace{3.3cm}
\begin{picture}(8,3.5)
\thicklines
%
%
\put(0.5,1){\line(1,0){2}}
\put(1.5,1){\vector(1,0){0}}
\put(0.8,1.3){$e_{Rj}$}
\put(2.5,1){\circle*{0.2}}
\put(2.4,0.4){$G_{jk}^*$}
\multiput(2.5,1)(0.5,0){8}{\line(1,0){0.3}}
\put(3.5,1){\vector(1,0){0}}
\put(3,1.3){$\phi$}
\put(4.5,1){\circle*{0.2}}
\put(5.5,1){\vector(1,0){0}}
\put(5.5,1.3){$\phi$}
\put(6.5,1){\circle*{0.2}}
\put(6.4,0.4){$G_{ik}$}
\put(6.5,1){\line(1,0){2}}
\put(8,1){\vector(1,0){0}}  
\put(8,1.3){$e_{Li}$}
\put(4.5,1){
\qbezier(-2,0)(-2.01,0.35)(-1.88,0.68)
\qbezier(-1.88,0.68)(-1.78,1.03)(-1.53,1.29)
\qbezier(-1.53,1.29)(-1.3,1.55)(-1,1.73)
\qbezier(-1,1.73)(-0.69,1.9)(-0.35,1.97)
\qbezier(-0.35,1.97)(0,2.03)(0.35,1.97)
\qbezier(2,0)(2.01,0.35)(1.88,0.68)
\qbezier(1.88,0.68)(1.78,1.03)(1.53,1.29)
\qbezier(1.53,1.29)(1.3,1.55)(1,1.73)
\qbezier(1,1.73)(0.69,1.9)(0.35,1.97)
}
\put(4.4,3){\circle*{0.2}}
\put(4.2,3.3){$(M_E)_{kk}$}
\put(2.2,2.3){$E_{Lk}$}
\put(6.6,2.3){$E_{Rk}$}
\end{picture}

\centerline
{\small\bf Fig.~1 Radiative mass generation of the charged leptons}

\vspace{3mm}
{\bf Suggestion (A)}:
The feature (a) suggests that the charged lepton 
mass spectrum is not originated in the Yukawa coupling 
structure at the tree level, 
but it is given by a bilinear form 
on the basis of a some mass generation mechanism. 
For example, 
in Refs. \cite{Koide82,Koide83}, 
a radiative-mass-generation-like mechanism 
shown in Fig.~1 has been assumed: 
$$
(M_e)_{ij} = m_0 \sum_{k} G_{ik} G^*_{jk} \ ,
\eqno(2.1)
$$
where $G_{ij}$ are coupling constants of the 
interactions 
$\overline{e}_i E_j \phi$.
On the other hand, in Refs.~\cite{Koide90, KF96, KT96, Koide99}, 
a seesaw-like mechanism \cite{UnivSeesaw} shown in Fig.~2 has 
been assumed: 
$$
M_e=mM^{-1}_E m^{\dagger}.
\eqno(2.2)
$$
In any cases, we need hypothetical heavy charged leptons $E$.

\vspace{-1.5cm}
\hspace*{4.5cm}
\unitlength=1pt
\begin{picture}(300,80)(50,50)
\put(0,50){\thicklines \vector(1,0){40}}
\put(40,50){\thicklines \line(1,0){35}}
\put(35,30){$e_{Li}$}
\put(75,50){\circle*{5}}
\put(70,60){$m_{ik}$}
\put(75,50){\thicklines \vector(1,0){40}}
\put(115,50){\thicklines \line(1,0){35}}
\put(115,30){$E_{Rk}$}
\put(150,50){\circle*{5}}
%
\put(150,50){\circle*{5}}
\put(145,60){$(M_E)_{k\ell}$}
\put(150,50){\thicklines \vector(1,0){40}}
\put(190,50){\thicklines \line(1,0){35}}
\put(185,30){$ E_{L\ell} $}
%
\put(225,50){\circle*{5}}
\put(220,60){$m^*_{j\ell} $}
\put(225,50){\thicklines \vector(1,0){40}}
\put(265,50){\thicklines \line(1,0){35}}
\put(265,30){$e_{Rj}$}
\end{picture}

\vspace*{1cm}
\centerline{\small\bf Fig.~2 Seesaw-like mass generation of $m_{ei}$}
\vspace{3mm}

{\bf Suggestion (B)}:
The feature (b) suggests that the theory is 
invariant under the permutation symmetry S$_3$.
As an example of the S$_3$ invariant mass matrix, 
the so-called democratic mass matrix \cite{democratic} 
is well known. 
The derivation of (1.2) in Ref.~\cite{Koide93} is 
based on a democratic mass matrix model. 
However, in the present paper, 
as we review in the next section, 
we will adopt another idea, where what is essential 
is not a structure of the Yukawa coupling constants, 
but a structure of the vacuum expectation values (VEVs) 
of flavor-triplet Higgs scalars. 

{\bf Suggestion (C)}:
The feature (c) suggests that the mechanism which leads 
to the relation (1.2) must work at a low energy scale. 
In the conventional model, the mass matrix structure 
is due to the Yukawa coupling structure, 
which is given at the unification energy scale 
$\mu= M_{GUT}$.
The mass spectrum at a low energy scale 
must be evaluated 
by taking the RGE effects into consideration. 
Against such the conventional models, the idea that mass spectrum 
is due to the VEV structure of Higgs scalars at a low energy scale
is very attractive as an explanation of the 
non-RGE-invariant mass formula. 

\vspace{5mm}
{\large\bf 3. \ S$_3$ symmetry and VEV of flavor-triplet-scalars}

The idea to relate the VEVs structure to 
a mass matrix model has first been proposed in 1990 
\cite{Koide90} (and also see \cite{KT96}) 
although the model was based not on the S$_3$ symmetry, 
but on a U(3) symmetry. 
A model based on the S$_3$ symmetry has been investigated 
in 1999 \cite{Koide99}.

The basic idea is as follows:
We consider the following S$_3$ invariant 
Higgs potential 
$$
V=\mu^2 \sum_{i}(\overline{\phi}_i \phi_i)
+\frac{1}{2} \lambda \left[\sum_i(\overline{\phi}_i \phi_i)\right]^2
+ \eta(\overline{\phi}_{\sigma} \phi_{\sigma})
(\overline{\phi}_{\pi}\phi_{\pi} + \overline{\phi}_{\eta}\phi_{\eta}), 
\eqno(3.1)
$$
where 
$(\overline{\phi}_i \phi_i)=\phi^-_i \phi^+_i 
+\overline{\phi} \ ^0_i \phi^0_i$ \ , and 
$$
\begin{array}{l}
\phi_{\pi} = \frac{1}{\sqrt{2}}(\phi_1 - \phi_2) \ , \\ 
\phi_{\eta} = \frac{1}{\sqrt{6}}(\phi_1 + \phi_2 -2\phi_3) \ , \\
\phi_{\sigma} = \frac{1}{\sqrt{3}}(\phi_1 + \phi_2 +\phi_3) \ .
\end{array}
\eqno(3.2)
$$
(For more general S$_3$-invariant Higgs potential, 
see Ref. \cite{Koide99}. Even under a more general 
S$_3$-invariant potential (but with some constraints), the relation 
(3.7) given below is unchanged.)
The conditions for the VEVs 
$v_i\equiv \langle \phi^0_i \rangle$ at which 
the potential (3.2) takes the minimum are 
$$
\mu^2 + \lambda \sum_i|v_i|^2 
+ \eta \left(|v_{\pi}|^2 + |v_{\eta}|^2 \right) =0 \ , 
\eqno(3.3)
$$
$$
\mu^2 + \lambda \sum_i|v_i|^2 + \eta |v_{\sigma}|^2 =0 \ , 
\eqno(3.4)
$$
so that we obtain
$$
|v_{\sigma}|^2= |v_{\pi}|^2 + |v_{\eta}|^2 
= \frac{-\mu^2}{2\lambda + \eta} \ .
\eqno(3.5)
$$
Therefore, from the relation 
$$
\overline{\phi}_1 \phi_1+ \overline{\phi}_2 \phi_2
+ \overline{\phi}_3 \phi_3 =
\overline{\phi}_\pi \phi_{\pi}+ \overline{\phi}_\eta \phi_{\eta}
+ \overline{\phi}_\sigma \phi_{\sigma} \ ,
\eqno(3.6)
$$
we obtain
$$
|v_1|^2 + |v_2|^2 + |v_3|^2 
=|v_{\pi}|^2 + |v_{\eta}|^2 + |v_{\sigma}|^2 
= 2|v_{\sigma}|^2
= 2 \left( \frac{v_1 + v_2 + v_3}{\sqrt{3}}\right) ^2.
\eqno(3.7)
$$
If we consider a model in which the charged lepton masses are 
given by 
$$
\sum_i \overline{e}_i \langle\overline{\phi}_i^0\rangle
\langle \phi_i^0 \rangle e_i \ ,
\eqno(3.8)
$$
we can obtain the charged lepton mass formula (1.2).
Note that the formula (1.2) can be derived 
independently of the values of $\lambda$ and $\eta$.
The relation (3.7) holds at the SU(2)$_L$ symmetry 
breaking energy scale $M_W$, 
the formula (1.2) is also valid at $\mu=M_W$. 

However, this scenario has some troubles. 
We know that the mass terms $m_{ei} (\bar{e}_{Li} e_{Ri}
+\bar{e}_{Ri} e_{Li})$ are $\Delta I=1/2$, while
Eq.~(3.8) will be come from $\Delta I =1$ terms.
Besides, in this model, there are 3-family Higgs scalars, 
so that they cause, in principle, 
flavor changing neutral currents (FCNC). 
Moreover, if we wish to build a GUT model, 
the 3 family scalars affect on the RGE effects dangerously, 
so that the beautiful unification of the gauge 
coupling constants, 
$g_1=g_2=g_3$ , at $\mu=M_{GUT}$ in the minimal 
SUSY GUT model will be spoiled. 

A most straightforward improvement 
of this model will be to change 
the 3-family SU(2)$_L$-doublet scalars $\phi_i$ into 
3-family SU(2)$_L$-singlet scalars $\phi^0_i$. 
When we introduce additional heavy fermions 
$(5'_L + 10'_L)_{(+)}$ and flavor-triplet SU(5)-singlet
scalars $\phi_{(-)}$ in addition to the quarks and
leptons $(\bar{5}_L +10_L)_{(-)}$ and Higgs fields
$\bar{H}_{d(-)} + H_{u(+)}$, where $(+)$ and $(-)$ 
denote the Z$_2$ charges, the charged lepton masses are 
given by a seesaw form 
$$
(M_e)_{ij} \simeq \delta_{ij}\langle \phi^0_i \rangle 
\langle H^0_u\rangle^{-1} 
\langle \phi^0_j\rangle
\eqno(3.9)
$$
as shown in Fig.~3. 
(Here, we have assumed the Z$_2$ symmetry in order forbid
the direct coupling of $\bar{H}_d$ to $\bar{5}_L 10_L$, so that
the leading terms are the seesaw mass terms given in (3.9).) 
However, if the additional fermions 
are indeed only $(5'_L + \overline{10}'_L)$, 
the fermions cannot be heavy, 
so that the 3-family fermions 
bring serious troubles into the theory. 
(For example, the color SU(3) does not become asymptotically free.)
Therefore, we must introduce further additional 
fermions $(\overline{5}'_L + 10'_L)$. 
However, then, the seesaw form (3.9) will be spoiled 
because of the dominant mass terms 
$(\mu_5\overline{5}'_L 5_L + \mu_{10}\overline{10}'_L 10_L)$. 

\vspace{-2cm}
\hspace*{4cm}
\begin{picture}(300,120)(50,50)
\put(0,50){\thicklines \vector(1,0){40}}
\put(40,50){\thicklines \line(1,0){35}}
\put(35,30){$\bar{5}_{L(-)}$}
\put(75,50){\thicklines \line(0,1){5}}
\put(75,60){\thicklines \line(0,1){5}}
\put(75,70){\thicklines \line(0,1){5}}
\put(75,80){\thicklines \line(0,1){5}}
\put(70,85){\thicklines \line(1,1){10}}
\put(80,85){\thicklines \line(-1,1){10}}
\put(75,50){\circle*{5}}
\put(65,105){$\langle \phi^0_{(-)} \rangle$}
\put(150,50){\thicklines \vector(-1,0){40}}
\put(110,50){\thicklines \line(-1,0){35}}
\put(115,30){$5'_{L(+)}$}
\put(150,50){\circle*{5}}
\put(150,50){\thicklines \line(0,1){5}}
\put(150,60){\thicklines \line(0,1){5}}
\put(150,70){\thicklines \line(0,1){5}}
\put(150,80){\thicklines \line(0,1){5}}
\put(145,85){\thicklines \line(1,1){10}}
\put(155,85){\thicklines \line(-1,1){10}}
\put(150,50){\circle*{5}}
\put(140,105){$\langle H^0_{u(+)} \rangle$}
\put(150,50){\thicklines \vector(1,0){40}}
\put(190,50){\thicklines \line(1,0){35}}
\put(185,30){$ \bar{10}'_{L(+)} $}
\put(225,50){\thicklines \line(0,1){5}}
\put(225,60){\thicklines \line(0,1){5}}
\put(225,70){\thicklines \line(0,1){5}}
\put(225,80){\thicklines \line(0,1){5}}
\put(220,85){\thicklines \line(1,1){10}}
\put(230,85){\thicklines \line(-1,1){10}}
\put(225,50){\circle*{5}}
\put(215,105){$\langle \phi^0_{(-)} \rangle$}
\put(300,50){\thicklines \vector(-1,0){40}}
\put(260,50){\thicklines \line(-1,0){35}}
\put(265,30){$10_{L(-)}$}
\end{picture}

\vspace*{1cm}
\begin{quotation}
{\small\bf Fig.~3 \ Seesaw mass $M_e$ in a model with 
fermions $3(\bar{5}_{L(-)} +10_{L(-)} +5'_{L(+)} +
\bar{10}'_{L(+)})$ and
scalars $(H_{u(+)} +3\phi^0_{(-)})$ under Z$_2$ symmetry
}
\end{quotation}
\vspace{3mm}

In the next section, 
we will propose a model under consideration 
of these problems. 

\vspace{5mm}
{\large\bf 4. Model}

According to the guiding principles (A), (B) and (C) 
suggested in Sec.~2 and the idea reviewed in Sec.~3, 
in this section, let us try to build a model 
which gives the formula (1.2) reasonably. 
In this section, we will concentrate our attention 
on the charged lepton masses, 
so that we will not touch the quark and neutrino masses. 
For convenience, we use notations and conventions 
in an SU(5) SUSY GUT model, 
but we do not always assume the SUSY GUT.

The basic idea in the present model is as follows:
we assume 3-family SU(5) singlet scalars 
instead of 3-family SU(2)$_L$ doublet Higgs scalars 
$\phi_i$ in Sec.~3. The SU(5) singlet fields 
do not cause FCNC, and do not affects the RGE effects 
of the gauge coupling constants. 

We assume the following flavor-triplet matter fields 
and flavor-singlet Higgs fields, 
$$ 
3(1_L + \overline{5}_L + 10_L)_{(+1)} 
+ 3(1'_L + \overline{5}'_L + \overline{5}'_L + 
\overline{10}'_L + 10'_L)_{(+2)} 
+ \overline{H}_{d(0)} + H_{u(+2)} \ , 
\eqno(4.1)
$$
where  $\overline{H}_d$ and $H_{u}$ denote SU(5) 
$\overline{5}$ and 5 Higgs fields, 
respectively, and the number $(n)$ 
in $\psi_{L(n)}$ denotes the Z$_{4}$ charge, 
i.e. $\psi_{L(n)} \rightarrow e^{i(\pi/2)n} \psi_{L(n)}$
under a discrete symmetry Z$_4$.
The Z$_4$ invariant superpotential $W$ is given by 
$$
W= (10_{Li}\ 10'_{Li}\ \overline{10}'_{Li})
\left(
\begin{array}{ccc}
Y^u_{ij} H_u & 0 & \lambda^u_{ijk} 1_{Lk} \\ 
0 & 0 & \mu_{10} \delta_{ij}  \\
\lambda^u_{ijk} 1_{Lk} & \mu_{10}\delta_{ij} & 
Y^{u\prime}_{ij} \overline{H}_d 
\end{array} \right)
\left(
\begin{array}{ccc}
10_{Lj} \\
10'_{Lj} \\
\overline{10}'_{Lj} 
\end{array}
\right)
$$
$$
\ \ \ \ \ \ \ \ \ \ \ \ \ \ 
+(\overline{5}_{Li}\ \overline{10}'_{Li}\ \overline{5}'_{Li})
\left(
\begin{array}{ccc}
0 & \lambda^{(5, 5')}_{ijk} 1_{Lk} & 0 \\
\lambda^{(10', 10)}_{ijk} 1_{Lk} & 0 & \mu_{10} \delta_{ij} \\
0 & \mu_{5} \delta_{ij} & Y^{d\prime}_{ij} \overline{H}_d 
\end{array} \right) \
\left(
\begin{array}{ccc}
10_{Lj} \\
5'_{Lj} \\
\overline{10}'_{Lj} 
\end{array} \right) \
$$
$$
\begin{array}{l}
\hspace{1.4cm}
+ \lambda^{(10, 10, 5')}_{ijk} 10_{Li} 10_{Lj} 5'_{Lk} \ 
+ \lambda^{(5, 5, 10')}_{ijk} \overline{5}_{Li} 
 \overline{5}_{Lj} 10'_{Lk} \\ 
\hspace{1.4cm}
+ \lambda^{(5', 5, 10)}_{ijk} \overline{5}'_{Li} 
 \overline{5}_{Lj} 10_{Lk} \ 
+ \lambda^{\nu}_{ijk} 1_{Li} 1_{Lj} 1'_{Lk} \\ 
\hspace{1.4cm}
+ Y^{\nu}_{ij} 1_{Li} \overline{5}_{Lj} H_u \ 
+ Y^{\nu\prime}_{ij} 1'_{Li} \overline{5}'_{Lj} \overline{H}_d 
+ Y^{(H)}_{i} 1'_{Li} \overline{H}_d H_u + \mu'_{5i} 
 \overline{5}'_{Li} H_u \ . 
\end{array}
\eqno(4.2)
$$

Although the up-quark masses are given by 
the Yukawa interactions 
$Y^u_{ij} 10_{Li} 10_{Lk} \langle H^0_u \rangle$ , 
the down-quarks and charged leptons do not 
have such Yukawa interactions at tree level, 
so that the mass matrices are given by the seesaw form 
$$
(M_{d,e})_{ij}\simeq \frac{1}{\mu_5\mu_{10}}
\lambda^{(5,5')}_{ii'k} \lambda^{(10',10)}_{jj'k'} 
 Y^d_{i'j'}v_{Sk}v_{Sk'}
\eqno(4.3)
$$
as shown in FIg.~4, where 
$v_{Si}= \langle 1_{Li} \rangle$.
We assume universality of the coupling constants
$$
\lambda^{(5,5')}_{ijk} = \lambda^{(10',10)}_{ijk}
\equiv \lambda \delta_{ij} \delta_{jk} \ , 
\eqno(4.4)
$$
$$
Y^d_{ij} = y_d \delta_{ij}.
\eqno(4.5)
$$
Then, we obtain a simple form 
$$
(M_{d,e})_{ij} = \delta_{ij}\lambda^2y_d 
\frac{v^2_{Si}v_d}{\mu_5 \mu_{10}} \ .
\eqno(4.6)
$$
Of course, for the scalar parts of the fields $1_{Li}$, 
we assume a similar mechanism to $\phi_i$ in the Higgs 
potential (3.3) as discussed in Sec.~3.
Then, we obtain the relation (1.2). 
(At present, we unwillingly obtain the same 
mass matrix form for the down-quarks, i.e. 
$M_d =M_e^T$.)

\vspace{-1cm}
\hspace*{4cm}
\begin{picture}(300,120)(50,50)
\put(0,50){\thicklines \vector(1,0){30}}
\put(30,50){\thicklines \line(1,0){20}}
\put(25,30){$\bar{5}_{L(+1)}$}
\put(50,50){\thicklines \line(0,1){5}}
\put(50,60){\thicklines \line(0,1){5}}
\put(50,70){\thicklines \line(0,1){5}}
\put(50,80){\thicklines \line(0,1){5}}
\put(45,85){\thicklines \line(1,1){10}}
\put(55,85){\thicklines \line(-1,1){10}}
\put(50,50){\circle*{5}}
\put(35,105){$\langle 1_{L(+1)}\rangle$}
\put(100,50){\thicklines \vector(-1,0){30}}
\put(70,50){\thicklines \line(-1,0){20}}
\put(75,30){$5'_{L(+2)}$}
\put(100,50){\circle*{5}}
\put(95,60){$\mu_5$}
\put(100,50){\thicklines \vector(1,0){30}}
\put(130,50){\thicklines \line(1,0){20}}
\put(125,30){$\bar{5}'_{L(+2)}$}
\put(150,50){\thicklines \line(0,1){5}}
\put(150,60){\thicklines \line(0,1){5}}
\put(150,70){\thicklines \line(0,1){5}}
\put(150,80){\thicklines \line(0,1){5}}
\put(145,85){\thicklines \line(1,1){10}}
\put(155,85){\thicklines \line(-1,1){10}}
\put(150,50){\circle*{5}}
\put(135,105){$\langle \bar{H}_{d(0)}\rangle$}
\put(200,50){\thicklines \vector(-1,0){30}}
\put(170,50){\thicklines \line(-1,0){20}}
\put(175,30){$10'_{L(+2)}$}
\put(200,50){\circle*{5}}
\put(195,60){$\mu_{10}$}
\put(200,50){\thicklines \vector(1,0){30}}
\put(230,50){\thicklines \line(1,0){20}}
\put(225,30){$\bar{10}'_{L(+2)}$}
\put(250,50){\thicklines \line(0,1){5}}
\put(250,60){\thicklines \line(0,1){5}}
\put(250,70){\thicklines \line(0,1){5}}
\put(250,80){\thicklines \line(0,1){5}}
\put(245,85){\thicklines \line(1,1){10}}
\put(255,85){\thicklines \line(-1,1){10}}
\put(250,50){\circle*{5}}
\put(235,105){$\langle 1_{L(+1)} \rangle$}
\put(300,50){\thicklines \vector(-1,0){30}}
\put(270,50){\thicklines \line(-1,0){20}}
\put(275,30){$10_{L(+1)}$}
\end{picture}

\vspace*{1cm}
\centerline{\small\bf Fig.~4 \  Charged lepton mass generation}

\vspace{4mm}

For the neutrino mass matrix $M_\nu$, we also obtain 
a seesaw form 
$$
(M_\nu)_{ij} \simeq  Y^{\nu}_{ii'}v_u
(\lambda^{\nu}_{i'j'k} v'_{Sk})^{-1}
Y^{\nu}_{jj'}v_u \ ,
\eqno(4.7)
$$
through the diagram given in Fig.~5, where 
$v'_{Si}= \langle 1'_{Li} \rangle$. 
Since $v_u \sim 10^2$ GeV, we suppose 
$v'_S \sim 10^{14}$ GeV as well as in the 
conventional seesaw model.

\vspace{-1cm}
\hspace*{4cm}
\begin{picture}(300,120)(50,50)
\put(0,50){\thicklines \vector(1,0){40}}
\put(40,50){\thicklines \line(1,0){35}}
\put(35,30){$\bar{5}_{L(+1)}$}
\put(75,50){\thicklines \line(0,1){5}}
\put(75,60){\thicklines \line(0,1){5}}
\put(75,70){\thicklines \line(0,1){5}}
\put(75,80){\thicklines \line(0,1){5}}
\put(70,85){\thicklines \line(1,1){10}}
\put(80,85){\thicklines \line(-1,1){10}}
\put(75,50){\circle*{5}}
\put(60,105){$\langle {H}_{u(+2)} \rangle$}
\put(150,50){\thicklines \vector(-1,0){40}}
\put(110,50){\thicklines \line(-1,0){35}}
\put(115,30){$1_{L(+1)}$}
\put(150,50){\circle*{5}}
\put(150,50){\thicklines \line(0,1){5}}
\put(150,60){\thicklines \line(0,1){5}}
\put(150,70){\thicklines \line(0,1){5}}
\put(150,80){\thicklines \line(0,1){5}}
\put(145,85){\thicklines \line(1,1){10}}
\put(155,85){\thicklines \line(-1,1){10}}
\put(150,50){\circle*{5}}
\put(135,105){$\langle 1'_{L(+2)} \rangle$}
\put(150,50){\thicklines \vector(1,0){40}}
\put(190,50){\thicklines \line(1,0){35}}
\put(185,30){$ 1_{L(+1)} $}
\put(225,50){\thicklines \line(0,1){5}}
\put(225,60){\thicklines \line(0,1){5}}
\put(225,70){\thicklines \line(0,1){5}}
\put(225,80){\thicklines \line(0,1){5}}
\put(220,85){\thicklines \line(1,1){10}}
\put(230,85){\thicklines \line(-1,1){10}}
\put(225,50){\circle*{5}}
\put(210,105){$\langle {H}_{u(+2)} \rangle$}
\put(300,50){\thicklines \vector(-1,0){40}}
\put(260,50){\thicklines \line(-1,0){35}}
\put(265,30){$\bar{5}_{L(+1)}$}
\end{picture}

\vspace*{1cm}
\centerline{\small\bf Fig.~5 Neutrino mass generation}
\vspace{3mm}

On the other hand, in order to obtain $M_e \sim$ 1GeV, 
we must suppose 
$$
v^2_S / \mu_5 \mu_{10} \sim 10^{-2} \ ,
\eqno(4.8)
$$
for $v_d \sim 10^2$ GeV. 
If we suppose $v_S \sim 10$ GeV, we have to take 
$\sqrt{\mu_5 \mu_{10}} \sim 10^2$ GeV. 
However, such a small value means that 
the additional matter fields of 6 families 
survive until $\mu = \sqrt{\mu_5 \mu_{10}}\sim 10^{-2} $ GeV, 
so that the asymptotic freedom 
of the color SU(3) is destroyed. 
Therefore, for example, 
we put $\mu_{10} \sim 10^{16}$ GeV and 
$\mu_{5} \sim 10^{14}$ GeV, so that we take 
$v_{S} \sim 10^{14}$ GeV.
Then, the matter fields affect the RGE effects as follows: 
$3+6$ families for $\mu > M_{GUT} \sim 10^{16}$ GeV; 
3 families of $(\overline{5}+10)_L$ 
and 6 families of $(\overline{5}' + 5')$ for 
$M_{GUT} > \mu \geq \mu_5 \sim 10^{14}$ GeV; 
3 families of $(\overline{5}+10)_L$ for 
$\mu_5> \mu \geq M_{weak}$.
Here, our parameter values are summarized as 
$$
\begin{array}{l}
M_{weak} \sim v_u \sim v_d \sim 10^2 \ {\rm GeV} , \\ 
\mu_{5} \sim v_S \sim v'_S \sim 10^{14} \  {\rm GeV} , \\
M_{GUT} \sim \mu_{10} \sim 10^{16} \ {\rm GeV} . \\
\end{array}
\eqno(4.9)
$$
For these parameter values, 
the gauge unification at $\mu=M_{GUT}$ is still kept 
as shown in Fig.~6.

\begin{center}

\epsfile{file=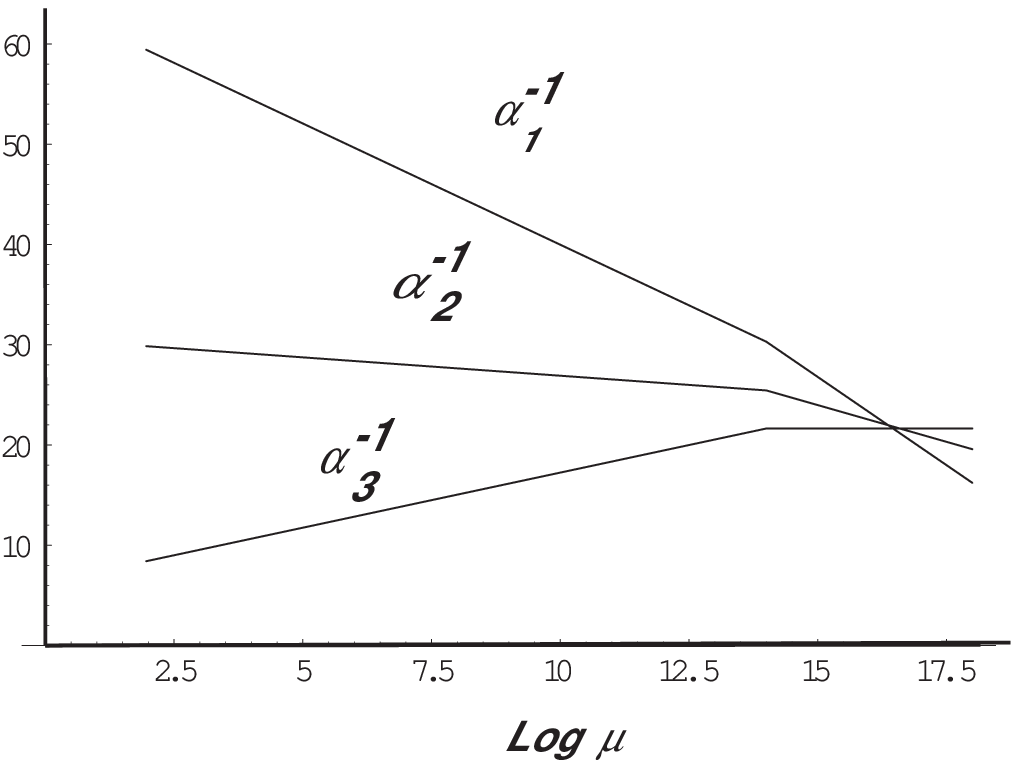,scale=0.9}

\begin{quotation}\small
{\bf Fig.~6  Evolution of the gauge coupling constants}
\end{quotation}
\end{center}

\vspace{5mm}
{\large\bf 5. \ Concluding remarks}

In conclusion, we have proposed a model 
which gives the charged lepton mass formula (1.2), 
where the origin of the mass spectrum is 
attributed not to the structure of the Yukawa coupling 
constants, but to a structure of VEVs 
of flavor-triplet scalars under $Z_4 \times S_3$ symmetries. 
The model can be described within a framework of the SUSY GUT.
(Of course, at present, the form of the scalar potential (3.1)
has been given by hand, it is not a logical consequence from
the SUSY GUT model.)

However, the choice $v_S \sim 10^{14}$ GeV does not 
satisfy the suggestion (C) that 
the formula (1.2) is valid at the law energy scale. 
If we adhere to the idea, 
we must give up the gauge unification. 
If we assume that only leptonic part of 
$(\overline{5}' + 5' + \overline{10}' + 10')_L$ 
survive untill 
$\mu \sim \mu_5 = \mu_5 \sim 10^2$ GeV 
(we assume a triplet-doublet splitting \cite{DTspl}
mechanism similar to that for the Higgs fields), 
we can choose $v_S \sim 10$ GeV without 
destroying the asymptotic freedom of the color SU(3).
In order to get the mass forumula (1.2) at a low enery
scale, rather, we should abandon the SUSY GUT senario.

So far, we have not discussed quark mass matrices 
(and also neutrino mass matrix). In the present model, 
the up-quark and neutrino mass matrices 
are generated by the diagrams given in Figs.~4 and 5, 
respectively. 
If we assume the universal couplings 
similar to (4.4) and (4.5), 
we will obtain diagonal forms of those 
mass matrices as well as in the charged lepton mass matrix, 
so that the Cabibbo-Kobayashi-Maskawa (CKM) 
and Maki-Nakagawa-Sakata (MNS) matrices will 
become unit matrices. 
Except for the charged lepton sector, 
we will have to consider a more general form 
of the coupling constants 
which is S$_3$ invariant. 
For example, we must consider that 
Yukawa couplings $Y^u_{ij}$ are given by the form 
$$
y^u_{(1)} 
\left( 
\begin{array}{ccc}
1 & 0 & 0 \\
0 & 1 & 0 \\
0 & 0 & 1 
\end{array} \right) \
+ y^u_{(2)}
\left(
\begin{array}{ccc}
0 & 1 & 1 \\
1 & 0 & 1 \\
1 & 1 & 0 
\end{array} \right) \ \ , 
\eqno(5.1)
$$
and the coupling constants $\lambda^{\nu}_{ijk} v'_{Sk}$ 
which come from the interactions 
$1_L1_L1'_L$ are given by 
$$
\lambda^{\nu}_{(1)}
\left(
\begin{array}{ccc}
v'_{S1} & 0 & 0 \\
0 & v'_{S2} & 0 \\
0 & 0 & v'_{S3}
\end{array}\right) \
+ 
\lambda^{\nu}_{(2)}
\left(
\begin{array}{ccc}
0 & v'_{S3} & v'_{S2} \\
v'_{S3} & 0 & v'_{S1}\\
v'_{S2} & v'_{S1} & 0 
\end{array}\right) \
\eqno(5.2)
$$
We must also consider a mechanism 
which yields $M_d \neq M^{T}_e$. 
Possibly, the mechanism will be related to the 
triplet-doublet splitting of 
the SU(5) $\overline{5}$ (and/or 5) fields. 

Thus, the present model has many problems, but
the formula (1.2) is too beautiful to be accidental
coincidence.
(Some of the problems will be solved by abandoning
the GUT scenario.)
In the present paper, we have investigated
a possible model within the framework of an extended seesaw
mechanism, but, on the other hand, the radiative mass generation 
hypothesis is also promising.
The idea that the origin of the
mass spectrum is attributed not to the structure
of the Yukawa coupling constants, but to the
structure of the VEVs of flavor-triplet scalars
will be worthwhile noticing.
It is a future task to seek for a more elegant and
simple model which can lead to the mass formula
(1.2).

\vspace{5mm}
\centerline{\bf Acknowledgment}

The author would like to thank Professor A.~ Rivero
for her interst in this topic and helpful comments
on this work.
He is also grateful to Professors J.~Kubo and 
D.~Suematsu, and to young Particle Physics members at 
Kanazawa University for their enjoyable and helpful
discussions. 


\end{document}